\documentclass[12pt]{article}

\usepackage{amsmath}         
\usepackage{amssymb}         
\usepackage{amsfonts}        
\usepackage{graphicx}        
\usepackage{slashed}

\usepackage{cite}          
\usepackage{mathrsfs}      

\usepackage[margin=2cm]{geometry} 

\newcommand{\arXiv}[2]{\href{http://arxiv.org/pdf/#1}{{\tt #2/#1}}}
\newcommand{\arXivold}[1]{\href{http://arxiv.org/pdf/#1}{{\tt #1}}}

\numberwithin{equation}{section}    

\renewcommand{\tilde}{\widetilde}

\newcommand{\beq}{\begin{eqnarray}}
\newcommand{\eeq}{\end{eqnarray}}

\newcommand{\drawsquare}[2]{\hbox{%
\rule{#2pt}{#1pt}\hskip-#2pt
\rule{#1pt}{#2pt}\hskip-#1pt
\rule[#1pt]{#1pt}{#2pt}}\rule[#1pt]{#2pt}{#2pt}\hskip-#2pt
\rule{#2pt}{#1pt}}

\newcommand{\Yfund}{\drawsquare{7}{0.6}}

\newenvironment{institutions}[1][2em]
  {\begin{list}{}{\setlength\leftmargin{#1}\setlength\rightmargin{#1}}\item[]}
  {\end{list}}

\usepackage[
	colorlinks=true,
	citecolor=black,
	linkcolor=black,
	urlcolor=blue,
	hypertexnames=false]{hyperref}

\begin{document}
\thispagestyle{empty}
\vspace{0.7 cm}
\begin{center}

	{	
		\LARGE \bf 
		Pre-ADS Superpotential from }\\
		\vskip .35cm
		{\LARGE \bf 
		Confined Monopoles
	}

	\vskip .7cm
	
	\renewcommand*{\thefootnote}{\fnsymbol{footnote}}

	{	
		\bf
		Csaba Cs\'aki$^{a}$, Mario Martone$^{b}$, Yuri Shirman$^{c}$,\\
		\vspace{.2cm}
		 and {John Terning}$^{d}$
	}  

	\vspace{.2cm}

\begin{institutions}[2.25cm]
\footnotesize

    
	$^{a}$ {\it Dept.\ of Physics, \textsc{lepp}, Cornell University, Ithaca, NY 14853}  
	\\
	$^{b}$ {\it Dept. of Physics, University of Texas, Austin, TX 78712}  
	\\
	$^{c}$ {\it Dept.\ of Physics \& Astronomy, University of California, Irvine, CA 92697} 
	\\
	$^{d}$ {\it Dept.\ of Physics, University of California, Davis, CA 95616} 
\end{institutions}	
\end{center}

\vspace{-.5em}
\begin{center}{ \footnotesize\tt csaki@cornell.edu, mariomartone@utexas.edu, yshirman@uci.edu,  jterning@gmail.com} 
\end{center}


\begin{abstract}
\noindent 
According to the standard lore only single monopoles contribute to the superpotential on the  Coulomb branch of 3D ${\cal N}=2$ SUSY gauge theories. However we argue that multi-monopole configurations can also generate superpotential terms in the presence of squark VEVs on the mixed Higgs-Coulomb branch. The new ingredient is the confinement of monopoles via Nielsen-Olesen flux tubes. Such confined multi-monopoles will yield a pre-ADS superpotential which depends both on the local Coulomb moduli and matter superfields but has no fractional powers. Once the lifted moduli are integrated out 
 the familiar ADS superpotential is obtained. Our results demonstrate the important role multi-monopoles can play in generating non-perturbative effects and also sheds light on the still somewhat mysterious dynamical origin of the general 4D ADS superpotential. 
\end{abstract}


\section{Introduction}
Exact non-perturbative results for SUSY QCD in 4D have played an important role in our understanding of the dynamics of strongly interacting gauge theories. One of the most famous examples of such dynamical effects is the non-perturbative  Affleck-Dine-Seiberg (ADS) superpotential \cite{Affleck:1983mk} when the number of flavors, $F$,  is less than the number of colors $N$. Surprisingly the exact dynamical origin of this term is still somewhat mysterious. For $F=N-1$ there is a reliable instanton calculation, however for fewer flavors 
the instanton generated correlation functions involve more than two fermions and thus can not be interpreted as a contribution to the superpotential. One can nevertheless deduce the presence of the ADS superpotential term via integrating out flavors, and attribute its dynamical origin to gaugino condensation in the unbroken $SU(F-N)$ subgroup. However no reliable direct calculation of this superpotential term currently exists. 

One promising approach \cite{Affleck:1982as,deBoer:1997kr,Aharony:1997bx,Lee:1998bb,Aharony:2013dha,Csaki:2014cwa,Poppitz:2013zqa} for shedding light on some of these dynamical effects is to compactify the 4D theory on a circle. Compactifying the theory significantly changes the structure both of the moduli space as well as the spectrum of non-perturbative objects in the theory. Due to the possible VEV of the component of the gauge field along the compact direction, the moduli space develops a Coulomb branch where the theory is generically described by a $U(1)^r$ gauge theory, where $r$ is the rank of the gauge group. Furthermore, out on the Coulomb branch there will be monopole-instantons present. These are the classical 't Hooft-Polyakov solutions which in 4D are interpreted as static monopoles, however when wrapped around the compactified direction they play the role of 3D instantons. They are more elementary objects than the 4D instanton, in fact the 4D instanton can be thought of as a multi-monopole containing one of each type of fundamental monopole \cite{Lee:1998bb}. It is believed that they can possibly contribute to more varied physical quantities than 4D instantons. For instance, it has been noticed in \cite{Aharony:1997bx} that gaugino condensation in a pure $SU(N)$ SYM theory  arises due the dynamics of $N$ individual monopoles of a theory on ${\bf R}^3\times {\bf S}^1$. Khoze et al.~\cite {Davies:1999uw}  explicitly calculated the resulting superpotential by combining the  effects of $N$ single monopole contributions and taking the large radius limit. They also argued that in theories with $0<F<N-1$ the superpotential can be obtained by going on a Higgs branch so that low energy physics is a pure SYM with $SU(N-F)$ gauge group and then performing a single monopole calculation in the low energy effective theory \cite{Davies:1999uh}. Our results will clarify this statement and confirm it in a somewhat modified form.

Specifically, in this paper we will examine the effects of multi-monopoles. Our key new insight is to point out that to correctly account for the monopole contribution to the superpotential, the calculation must be done in a specific region of the moduli space, in particular on a mixed branch where both Coulomb and squark VEVs can be turned on. The effect of the latter is to break some of the $U(1)$'s and confine some of the fundamental monopoles via a Nielsen-Olesen string \cite{Nielsen:1973cs,Nambu:1977ag} (aka a magnetic flux tube \cite{`tHooftMandelstam}). We will argue that in 3D such confined multi-monopole configurations actually contribute to the superpotential. We find a pre-ADS superpotential of the form 
\beq
\label{eq:preADS}
W_{\rm pre-ADS} = \frac{1}{\Pi_{i=1}^{F+1} Y_i \, {\rm det} Q\overline{Q}} 
\label{eq:preADSstart}
\eeq 
where the $Y_i$'s are the local Coulomb moduli corresponding to the monopoles that are confined by the squark VEVs. We will show that for $SU(N)$ with $F$ flavors a multi-monopole made up of $(F+1)$ confined fundamental monopoles has exactly the right number of zero modes and quantum numbers to contribute to this pre-ADS superpotential, and we present a simple accounting of the relevant contributions to the path integral showing that (\ref{eq:preADS}) is indeed the right form of the resulting pre-ADS superpotential. The ordinary, single monopole contributions of Affleck, Harvey and Witten (AHW)~\cite{Affleck:1982as} will still be present for the Coulomb moduli associated with unconfined monopoles. Eliminating the lifted moduli will give rise to the 3D ADS superpotential first described in~\cite{Aharony:1997bx}, while taking the $R\to \infty$ limit we obtain the usual form of the 4D ADS superpotential. Our results suggest that the true dynamical origin of the ADS superpotential lies in the confined multi-monopole contributions to the path integral. 

 Another approach to three dimensional physics that has received attention in recent years is 3D bosonization \cite{Giombi:2011kc} and the corresponding   flows of 
 relevant ${\mathcal N}=2$ theories \cite{Gur-Ari:2015pca}. While our results are not directly along this line of research, we hope that they could nevertheless shed light into a deeper understanding of the connections between 3D and 4D physics with the hope of contributing to progress towards finding analogs of bosonization in 4D.

The paper is organized as follows. Section \ref{sec:preADSstart} contains an overview of the results obtained in this paper and its relation to the already established web of non-perturbative results in 3D and 4D. In Section \ref{sec:SU3} we use symmetries and zero mode counting to obtain multi-monopole contributions to pre-ADS superpotential in $SU(3)$ theory with one flavor.
In Section \ref{sec:pathintegral} we show how this same result emerges from the path integral calculation of fermion correlation functions. In Section \ref{sec:SUN} we generalize the result for arbitrary $SU(N)$ with $F<N$.
Finally we present our conclusions in Section \ref{sec:conclude}. We close the paper with a series of Appendices. Appendix~\ref{app:monopolesolution} contains a review of the 't Hooft-Polyakov monopole embedded into $SU(3)$. Appendix~\ref{app:roots} reviews roots for $su(N)$ Lie algebras. Appendix~\ref{app:3DSUSY} has a general review of the essential elements of ${\cal N}=2$ 3D SUSY, while Appendix~\ref{app:KK} contrasts the properties of the pure 3D theory with those of the theory on $R^3\times S^1$. Finally, Appendix~\ref{app:Effective} contains a detailed description of multi-monopole zero modes.


\section{The Pre-ADS Superpotential}
\label{sec:preADSstart}
We will start out by giving a brief overview of our main result and its connection to other established non-perturbative effects. In the following sections we will argue that on the Coulomb branch of $SU(N)$ SUSY gauge theories with $F$ flavors on ${\bf R}^3\times {\bf S}^1$ there are important confined multi-monopole effects in the presence of squark VEVs. For $F<N$ one can go to a region of the Coulomb branch\footnote{Indeed, non-perturbative single monopole contributions have the effect of pushing the theory towards the region of the Coulomb branch where Higgs branch is accessible too \cite{Aharony:1997bx}.} where all squarks obtain VEVs.  These VEVs will break $F$ of the original $N-1$ $U(1)$ gauge symmetries and lead to  confinement of  $F+1$ fundamental monopoles. 
One of these monopoles has $2$ gaugino and $2F$ fundamental zero modes,\footnote{We assume for simplicity that all real mass terms vanish.} while the remaining fundamental monopoles have two gaugino zero modes each. 
The squark VEVs will lift all but two zero modes of the multi-monopole due to the supersymmetric Yukawa coupling mixing the gauginos and quarks.  Note that in supersymmetric theories there is an intricate relation between confinement, lifting of the zero modes and higgsing the $U(1)$'s. By supersymmetry the lifting of a gaugino zero mode must be accompanied by the lifting of the corresponding gauge boson mass, and hence the higgsing of a gauge symmetry.  
On a Coulomb branch this means that one of the $U(1)$'s must be broken in order for each mixing of a photino and quark to be allowed. However as 't Hooft and Mandelstam explained~\cite{`tHooftMandelstam}, an electrically charged VEV leads to the confinement of magnetic charges.  
Thus for the case of $F$ flavors with maximal rank of the matrix of VEVs, $F$ $U(1)$ factors will be broken, confining $F+1$ monopoles, and the resulting confined multi-monopole will have just two remaining unlifted fermionic zero modes, which will contribute to the superpotential and result in the pre-ADS term of (\ref{eq:preADS}). In addition single monopoles in the remaining unbroken $U(1)$'s will generate corresponding AHW $1/Y_i$ superpotential terms~\cite{Affleck:1982as}. Finally, since we are considering the theory on a circle, the KK monopole \cite{Lee:1997vp} will have its own $\eta \Pi_i Y_i$ contribution\footnote{Here we follow the notation of \cite{Aharony:1997bx}, so $\eta=\exp(-8\pi^2/g^2)\propto\Lambda^{b}$.}. The full superpotential in one of these regions of the Coulomb branch can be written as 
\beq
W_{pre-ADS}= \eta \Pi_i Y_i + \frac{1}{{\rm det}\,Q \overline{Q}\,\Pi_{i=1}^{F+1}Y_i } +\sum_{i=F+2}^{N-1} \frac{1}{Y_i}~.
\eeq
As usual local Coulomb moduli are lifted here, and can be integrated out of the theory. This can be conveniently done by introducing the globally defined Coulomb branch modulus $Y$ by adding  a Lagrange multiplier term $\lambda (Y - \Pi_i Y_i)$ to the superpotential. Once the local Coulomb branch moduli are integrated out we obtain the globally defined 3D ADS superpotential term of~\cite{Aharony:1997bx}:
\beq
\label{eq:W3D}
W_{3D} = \eta Y + (N-F-1) \frac{1}{(Y\,{\rm det}Q {\overline Q})^{\frac{1}{N-F-1}}}\,.
\eeq
While $Y$ is globally defined, it is easy to see that it is also lifted by the superpotential (\ref{eq:W3D}). Integrating out this last Coulomb branch modulus removes any obstacle to taking $R\to\infty$ limit. As expected we obtain the usual ADS superpotential valid both in a theory on a circle and in 4D limit superpotential:
\beq
\label{eq:WADS1}
W_{4D}=(N-F)\left(\frac{\Lambda^{3N-F}}{\det Q\overline{ Q}}\right)^\frac{1}{N-F}~.
\eeq

\section{Confined Monopoles in $SU(3)$ and the pre-ADS Superpotential}
\label{sec:SU3}
We begin the analysis by considering an $SU(3)$ SUSY gauge theory with a single flavor, $F=1$. This is the simplest interesting example where the superpotential does not arise either from single monopole contributions of the theory on a circle (which would require $F=0$) or from the instanton contributions in 4D limit (which would require $F=2$). The discussion of this section is easily generalized to an $SU(N)$ theory with one flavor, while the generalization to an arbitrary number of flavors will be considered in Sec.~\ref{sec:SUN}. We will show that in the ${\bf R}^3\times {\bf S}^1$ theory with $F=1$ two-monopole configurations will give rise to the pre-ADS superpotential: a superpotential involving inverse powers of the fields but no fractional powers. Since it is generated on regions of the Coulomb branch where also some squark VEVs are turned on it will contain both the matter fields as well as the Coulomb moduli.

For the $SU(3)$ gauge group there are two fundamental monopoles described by the Coulomb moduli $Y_{1,2}$ as well as a KK monopole (see Appendix~\ref{app:KK} for a review). 
We first consider the pure Coulomb branch of the theory, with no squark VEVs turned on.  
We will work in the fundamental Weyl chamber
\beq
\phi={\rm diag}(v_1,v_2,v_3)
\label{eq:123VEV}
\eeq
where $v_1 > v_2>v_3$ and $v_1 + v_2+v_3=0$.
Further assuming that $v_1>0>v_2$ we find the standard one-instanton superpotential terms on the Coulomb branch
\beq
W= \eta Y_1 Y_2 +\frac{1}{Y_2}
\eeq
The first term arises from the KK monopole while the second term is from the second fundamental monopole.  Both of these monopoles have two gaugino zero modes, and hence contribute to the gaugino two point function. For the first fundamental monopole however there are also  zero modes corresponding to $Q$ and $\overline{Q}$ in addition to the gaugino zero modes, so this monopole contributes to a fermion four point function rather than a two point function, and thus this monopole does  not contribute to the superpotential.  

Now consider two-monopole configurations on the mixed Coulomb-Higgs branch with non-vanishing squark VEVs $\langle Q\rangle$,  $\langle {\overline Q}\rangle$.  In particular we consider the following region: 
\beq
\phi={\rm diag}(v,0,-v) , \ \ \langle Q\rangle= \left( \begin{array}{c} 0 \\ q \\ 0 \end{array} \right) , \ \ \langle \bar{Q} \rangle = \left( \begin{array}{c} 0 \\ \bar{q} \\ 0 \end{array} \right) , \ \ |q|=|\bar{q}|\ .
\label{coulombsubbranch}
\eeq
While the adjoint VEV still breaks $SU(3)$ to $U(1)_1\times U(1)_2$, the squark VEVs further break the gauge symmetry: $U(1)_1\times U(1)_2\rightarrow U(1)_Q$, with the unbroken charge generator given by 
\beq
Q_{1+2}=\frac{1}{2}\,{\rm diag}(1,0,-1) ~,
\label{diaggen}
\eeq
and the broken generator given by
\beq
X=\frac{1}{2\sqrt{3}}\,{\rm diag}(1,-2,1) ~.
\label{brokengen}
\eeq
It's the dynamics of this additional breaking that underlies the generation of the pre-ADS superpotential. 
In 4D this breaking would confine \cite{`tHooftMandelstam,Nambu:1977ag} the first and second monopole by a Nielsen-Olesen vortex \cite{Nielsen:1973cs} to form a composite monopole that is only charged under the unbroken $U(1)_Q$. 
After Wick rotations and compactification the two fundamental monopoles are confined in 3D spacetime. This composite monopole potentially has 4 gaugino zero modes and 2 quark zero modes. However, as we shall see momentarily, 4 out of 6 zero modes are lifted by the squark VEVs and our composite monopole will generate a new superpotential term. At the same time, the $1/Y_2$ contribution of the second fundamental monopole disappears since, in the presence of the squark VEVs a single confined, fundamental monopole has an infinite length flux tube attached to it, and hence an infinite action.

\subsection{Zero mode structure of the composite monopole}

The most important ingredient needed to explain the origin of the superpotential term is an understanding of the zero mode structure of the composite monopole. As we saw above, the classical two-monopole configuration potentially has 4 gaugino and 2 quark zero modes while only configurations with two fermionic zero modes will contribute to the superpotential. Thus naively the two-monopole configuration could not contribute. However, this naive expectation is modified in the presence of squark VEVs in two important ways:
\begin{itemize}
\item in the presence of the squark VEVs the SUSY Yukawa coupling will lift two gaugino and two quark zero modes;
\item once the two monopoles are separated by some distance, $\rho \neq 0$, the remaining two exact zero will be mixed with the quark field, and thus the exact zero mode will live partly in the gauginos and partly in the quarks.
\end{itemize}
Let us discuss these statements in more detail. Irrespective of the presence of squark VEVs there are always two supersymmetric zero modes which can be obtained by a SUSY transformation on the monopole fields:
\beq
\label{eq:4Dlambda0}
\lambda_\alpha^{(0)}\propto \left(\sigma^{\mu\nu}\right)^\beta_{\alpha}\,F_{\mu\nu}\epsilon_\beta~.
\eeq
where $\epsilon$ is the supersymmetry transformation parameter. These will be the gaugino components of the exact zero modes, and as we will see shortly their nature depends on the separation, $\rho$, between the monopoles. First consider the $\rho=0$ case when the two monopoles are exactly on top of each other. In this case the two monopoles form a single monopole charged under $U(1)_Q$.  For this single monopole we can use the Callias index theorem \cite{Callias:1977kg,deBoer:1997kr} to confirm that we have four gaugino and two fundamental fermionic zero modes. Carefully analyzing how the gaugino zero modes transform under the full $SU(3)$, we can explicitly check that two of them are lifted in the presence of the squark VEVs due to the Yukawa coupling 
\begin{equation}
\sqrt{2}\,g\, Q^{*i} \lambda^j_i \psi_{j}\ .
\label{susyyukawa}
\end{equation}
Indeed, turning on the middle color component of the squark VEV, $q_{2}$, we get mass terms $\lambda^j_2 \psi_j$ and ${\overline \psi}^{\,i}\lambda^2_i$ which connect the two charged gaugino zero modes with the matter zero modes. Thus for the $\rho =0 $ case in the presence of squark VEVs the two unlifted zero modes purely live in the $U(1)_Q$ gauginos. The situation changes when $\rho \neq 0$: while four out of six zero modes are still lifted by the squark VEVs, the nature of two remaining zero modes is different. In this case,  instead of describing a pure monopole in the $U(1)_{Q}$ direction, we really need to think of it as a confined multi-monopole configuration. The main new physical effect is that the true zero modes are given by a mixture of $U(1)_Q$ and $U(1)_X$ gauginos. This is because when the monopoles are separated, with $g |q| \ll 1/\rho$, there is a dipole field corresponding  to the $U(1)_X$ charge (see Appendix~\ref{app:Effective} for a detailed discussion). Most importantly, the $\lambda_2^2$  component of the gaugino zero mode will no longer vanish. This means that the squark VEV $\langle Q \rangle =(0,q,0)$  induces a non-vanishing mixing, via (\ref{susyyukawa}), between the anti-chiral gaugino zero modes and (formerly non-zero mode) quarks. Thus for $\rho \neq 0$ we find that the remaining unlifted zero mode has a small admixture of the quark fields, and this admixture of the quark component in the exact zero mode will be proportional to $gq^*\rho$. We see that this multi-monopole configuration has exactly 2 zero modes which are partly contained in the quarks and partly in the gauginos. Hence they can contribute to a superpotential term of the form $1/(Y_1Y_2 Q\bar{Q})$.

\subsection{Small squark VEVs}

\begin{figure}
\begin{center}
\includegraphics[width=0.45\textwidth]{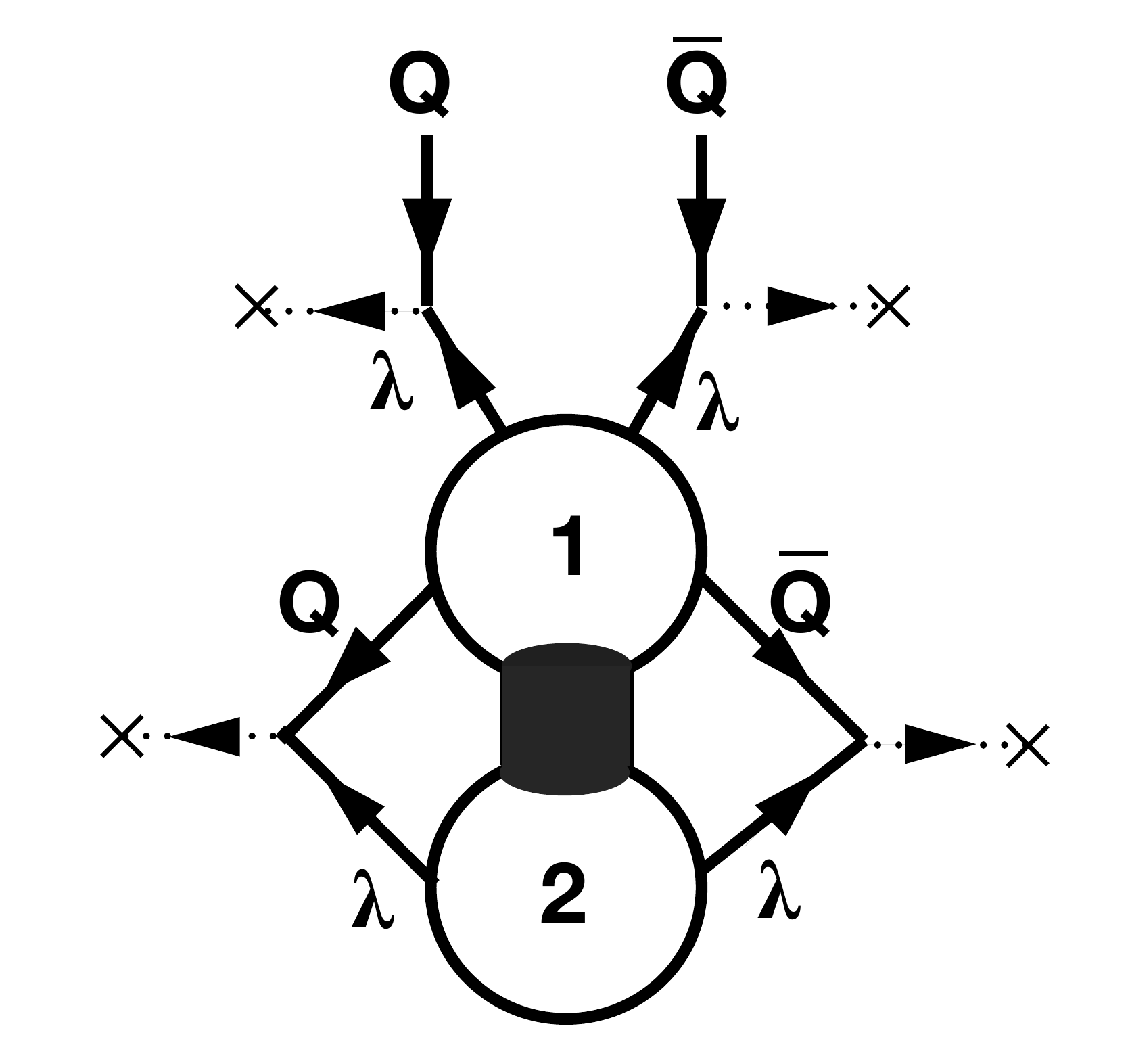}
\end{center}
\caption{Sketch of the multi-monopole contribution to the superpotential for $SU(3)$ with a single flavor. The dark cylinder connecting the two monopoles represents the flux tube confining the two fundamental monopoles.\label{fig:multiinst}}
\end{figure}

Let us first consider the case where $q,\overline{q}\ll v$. We will argue that in this case the superpotential is given by
\beq
W= \eta Y_1 Y_2 +\frac{1}{Y_1 Y_2 Q \overline{Q}}= \eta Y +\frac{1}{YQ{\overline Q}}\ 
\label{eq:effectiveW}
\eeq
where $Y=Y_1 Y_2$ is the globally defined Coulomb modulus.
The first term is still the contribution of the KK monopole. However the second term is the result of the first and second BPS monopoles being confined to form a single composite monopole.  As explained above, naively the confined monopole has too many zero modes to be able to contribute to the superpotential, but in the presence of the squark VEVs some of these zero modes are actually lifted. In a diagrammatic language this would corresponding to the closing up of the zero mode legs (``soaking up the zero modes" in the language of ADS). We illustrate the closing up of the zero modes of the confined multi-monopole in Fig.~\ref{fig:multiinst}. Note that in the presence of squark VEVs the remaining true zero modes contain a mixture of the anti-chiral gaugino and chiral matter fields where the mixing is proportional to $gq^*\rho$ (see the structure of the external fermion legs in Fig.~\ref{fig:multiinst}).

The more precise statement is that in the presence of the squark VEVs the multi-monopole configuration also contributes to the matter 2-point function, and the resulting contribution is holomorphic, hence corresponding to an effective superpotential term. The precise form of the resulting contribution to the 2-point function will determine the exact form of the superpotential. We will present a detailed analysis of the  multi-monopole contributions to the path integral evaluation of the two point correlators in the next section. Here we will only use symmetry considerations to restrict the possible form of a superpotential term. The $SU(3)$ SUSY gauge theory with one flavor has three $U(1)$ symmetries (one of them is anomalous). We can assign charges to the Coulomb moduli $Y_i$ under these symmetries, which match the charges of the zero modes for the corresponding monopole. Since we are considering the confined multi-monopole, the proper Coulomb modulus here will be $Y=Y_1 Y_2$, which is automatically the globally defined modulus (for general values of $F$ and $N$ this is only true when $F=N-2$). The assignment of charges is given by 
\beq
\begin{array}{|c|cccc|}  \hline 
& SU(3) & U(1)_A & U(1)_B & U(1)_R \\ \hline 
Q & \Yfund & 1 & 1 & 0 \\
\bar{Q} & \overline{\Yfund} & 1 & -1 & 0 \\
Y=Y_1Y_2 & 1 & -2 & 0 & -2 \\
\hline 
\end{array}
\eeq
We can see that the only superpotential term allowed by the symmetries is indeed given by \cite{Aharony:1997bx}
\begin{equation}
W_{\rm pre-ADS}= \frac{1}{Y Q\bar{Q}}\ .
\end{equation}
Symmetries allow this term, we have outlined the dynamical mechanism for generating it via confined multi-monopoles, and in the next section we will see that it is indeed generated with a non-zero coefficient. While we will not directly calculate the overall coefficient, consistency with known non-perturbative results requires it to be 1 for our specific case of $N=3, F=1$.
Adding on the $\eta Y$ term from the KK monopole one can integrate out the remaining Coulomb branch modulus $Y$ and then take $R\to \infty$ limit to find the expected 4D ADS superpotential \cite{Affleck:1983mk} 
\beq
W=2\left(\frac{\eta}{\det Q{\overline Q}}\right)^\frac{1}{2} =2\left(\frac{\Lambda^{8}}{\det Q{\overline Q}}\right)^\frac{1}{2}\,.
\eeq

\subsection{Large squark VEVs}

Now let us consider the opposite case with large $q, \overline{q}\gg v$.  The calculation is further simplified by taking $q, \overline{q} \gg  1/R$. In this case the gauge group is broken in the 4D regime so we first match the $SU(3)$ gauge theory with one flavor onto the low-energy effective theory with $SU(2)$ and no flavors:
\beq
\frac{1}{g_{L3}^2}&=&\frac{2\pi R}{g_{L4}^2(1/R)}=2\pi R\left[\frac{1}{g^2_{4}(\mu_0)} +\frac{b_L}{16\pi^2} \ln\left( \frac{1}{R^2\,Q\overline{Q}}\right) +\frac{b}{16\pi^2} \ln \left(\frac{Q\overline{Q}}{\mu^2_0}\right)\right]\\
\frac{4\pi}{R\,g_{L3}^2}&=&\frac{4\pi}{R\,g_{3}^2}+\frac{(b-b_L)}{2}\ln \left(R^2 Q{\overline Q}\right)~.
\eeq
In terms of the strong coupling scale we have
\beq
\label{eq:matching}
\Lambda^b=\Lambda^8 = \Lambda_L^6 Q{\overline Q}=\Lambda_L^{b_L} Q{\overline Q}~.
\eeq

On ${\bf R}^3\times {\bf S}^1$ the superpotential of the pure SYM $SU(2)$ is generated by contributions of single fundamental and singe KK monopoles:
\beq
W=\eta_L Y_L +\frac{1}{Y_L}~.
\eeq
Using (\ref{eq:matching}) we find $\eta=\eta_L Q\overline{Q}R^2$. Moreover, semiclassically the contribution of the KK monopole is independent of the squark VEVs, $\eta_LY_L=\eta Y$. We can now relate the Coulomb branch moduli of low and high energy physics: 
\beq
Y_L=YQ\overline{Q}R^2=Y_1Y_2Q\overline{Q}R^2~. 
\eeq
We conclude that the superpotential obtained in the large squark VEV regime agrees with the superpotential of obtained in the small squark VEV regime.


\section{Multi-Monopole Contributions to the Path Integral}
\label{sec:pathintegral}

In this section we will calculate contributions of two-monopole configurations to the superpotential of $SU(3)$ SUSY QCD with one flavor on ${\bf R}^3\times {\bf S}^1$ via the path integral. Our results can be generalized to $(F+1)$-monopole calculations in theories with $F$ flavors. These calculations are analogous to the calculation of constrained instantons which generate a superpotential  in 4D theories with $F=N-1$ flavors \cite{Affleck:1983mk}. Indeed, $N$-monopole configurations correspond to periodic instantons \cite{Harrington:1978ve} of ${\bf R}^3\times {\bf S}^1$ theories and turn into the usual 4D instantons in the large radius limit \cite{Lee:1998bb}. Thus it is useful to first briefly review the 4D instanton calculation \cite{Affleck:1983mk,Cordes:1985um} before attacking the case of confined multi-monopoles.

\subsection{Review of the 4D instanton calculation for generating the ADS superpotential for $F=N-1$}

In 4D ${\cal N}=1$ $SU(N)$ SUSY QCD with $F$ flavors there are $2N$ gaugino and $2F$ fundamental fermion zero modes in the one-instanton background. In the presence of the most generic squark VEVs all but two fermionic zero modes are lifted when $F=N-1$. The remaining two zero modes lead to a non-trivial instanton generated two-point chiral correlation function. The existence of a two fermion chiral correlation function implies an effective fermion mass and ADS superpotential in the low energy theory. The explicit evaluation of the two point correlation function requires knowledge of several factors: the classical action of the constrained instanton in the presence of squark VEVs; the contribution of bosonic and fermionic zero modes to the path integral\footnote{Due to supersymmetry contributions of non-zero modes in 4D cancel between bosons and fermions \cite{DAdda:1977sqj}.} and the mixing between exact zero modes with matter fermions. We will summarize the calculation of these factors below.

Up to a gauge transformation the gauge field of an instanton centered at the origin is given by 
\beq
 A_\mu^a=2\bar\eta_{a\mu\nu}\frac{x^\nu}{x^2+\rho^2}~,
\eeq
where $\rho$ is the instanton size and $\eta_{a\mu\nu}$ are the 't Hooft symbols and the instanton is embedded in the $SU(2)$ subgroup generated by $T^a=\tau^a$, $a=1,2,3$. In the presence of squark VEVs only the zero size instanton extremizes the action, thus it is necessary to use the constrained instanton formalism of ref.~\cite{Affleck:1983mk}. The squark profile in the fixed instanton background is given by
\beq
Q_{jf}=i \frac{(x^2)^{1/2}}{\left(x^2+\rho^2\right)^{1/2} }\,q\,\delta_{jf}~,~~~~~{\overline Q}^*_{jf}=i\frac{(x^2)^{1/2}}{\left(x^2+\rho^2\right)^{1/2} }\,\overline q^*\,\delta_{jf}~,~~~~~|q|=|\overline q|~.
\label{scalarprofile}
\eeq
The classical action of this field configuration is
\beq
 S(\rho)=\frac{8\pi^2}{g^2}+4\pi^2\rho^2|q|^2~.
\eeq

The instanton has $4N$ bosonic zero modes. Of these, $4$ correspond to spacetime translations of the instanton, $1$ to dilatations and $4N-5$ modes correspond to global rotations of the instanton in $SU(N)$. Naively, the existence of zero modes leads to divergences in the path integral. To obtain a physical result one must regulate the theory and integrate over collective coordinates corresponding to the location and size of the instanton. After calculating the corresponding Jacobian one finds that the measure of integration over the bosonic collective coordinates is
\beq
 d\mu_B\propto \frac{d^4x_0 d\rho}{\rho^5}\left(\frac{\rho\mu_0}{g}\right)^{4N}~,
\eeq
where $\mu_0$ is the regulator mass. Notice that this expression contains a factor of $\rho\mu_0/g$ for each bosonic zero mode. The dependence on the regulator mass $\mu_0$ reflects the contributions of the Pauli-Villars regulator fields (more specifically, the contributions of their lowest eigenvalues) to the path integral, while the dependence on $g$ arises from the Jacobian. Originally 't Hooft \cite{tHooft:1976snw} determined the dependence of the measure on the instanton size $\rho$ by using dimensional analysis; however, this dependence can also be obtained by carefully including the norm of the zero modes in the calculation of the Jacobian \cite{Bernard:1979qt}.

In the calculation of the fermionic contribution to the correlation function we must remember that the non-trivial scalar profiles (\ref{scalarprofile}) perturb the fermionic zero modes. In particular, all but two gaugino zero modes are lifted by the squark VEVs. Since supersymmetry requires the existence of Yukawa couplings 
\beq
{\mathcal L}_Y=\sqrt{2}g \lambda \psi q^*+\sqrt{2}g \lambda {\overline \psi} {\overline q}^*+{\rm h.c.} ~,
\label{yukawa}
\eeq
each squark VEV  $q^*$ or $\overline q^*$ soaks up one gaugino and one quark zero mode (cf. the closed fermion lines in Fig.~\ref{fig:multiinst}).
Thus in the presence of squark VEVs the lowest eigenvalues of these fields in the instanton background are lifted to $\sqrt{2}gq^*$. Correspondingly, contributions from the Pauli-Villars regulators to the path integral give a factor of $\mu_0^{-1}$ for each connected pair of fermion zero modes. To construct the integration measure for fermionic zero modes one must further include the contribution of the two surviving exact zero modes. This contribution includes the Grassmannian differentials of the exact fermionic zero modes, $d\xi$ and $d\bar \xi$, and a factor of $1/\mu_0$ arising from the corresponding regulator fields. The final expression for the fermionic measure with $F=N-1$ takes the form
\beq
 d\mu_F\propto d\xi d\bar\xi\left(\frac{g^2q^*\overline{q}^*}{\mu_0^2}\right)^{N-1}\frac{1}{\mu_0}\,.
 \label{fermionzeromodemeasure}
\eeq

The two exact zero modes are mostly gaugino zero modes (\ref{eq:4Dlambda0}), where, by supersymmetry,  the profile of the gaugino exactly follows the profile of the field strength. 
While these components of the exact zero modes lead to a gaugino-gaugino correlation function, it is of no interest in 4D due to the $1/x^4$ fall off of the zero modes. On the other hand, analogous components of zero modes will be important to us later when studying theories on a circle. In addition, the squark VEVs mix the superconformal gaugino zero modes with anti-chiral quark (and anti-quark) fields which contribute another piece of the exact zero mode. This can be seen from equations of motion for anti-chiral matter fermions which,  to leading order in $ g q^*\rho$, take the form \cite{Affleck:1983mk,Cordes:1985um}
\beq
 \slashed{D}\lambda&=&0\\
 \label{antichiraleom}
 \slashed{D}^\dagger\psi^{\dagger[\beta]}_{\alpha}&=&\sqrt{2}g\,Q^*\lambda^{0}\\
  \label{antichiraleom2}
\slashed{D}^\dagger\overline{\psi}^{\dagger[\beta]}_{\alpha}&=&-\sqrt{2}g\,\overline{Q}^*\lambda^{0} \,.
\eeq
The solution of the first of these equations is given by (\ref{eq:4Dlambda0}). Using explicit solutions for $Q(x)$ and $\lambda^0(x)$ it is easy to see that the quark and anti-quark components of the exact zero modes fall of as $1/x^3$ at large distance. In fact, an exact solution can be found by observing that, by supersymmetry, the righthand-sides of (\ref{antichiraleom}) and  (\ref{antichiraleom2}) are related to the derivative of the scalar profile \cite{Affleck:1983mk,Cordes:1985um}. To leading order in $gq^*\rho$ one finds that the component of the zero mode, $\chi$, is
\begin{equation}
 \chi^{[\beta]}_\alpha(x)\propto gq^* \slashed\partial_{\alpha}^{\beta}\left(\frac{(x-x_0)^2}{(x-x_0)^2+\rho^2}\right)^{1/2}\xi\propto gq^*\rho^2 S_4(x-x_0)\xi \, ,
\end{equation}
where $S_4$ is the 4D position space fermionic propagator. 
While the $\rho$ dependence in this expression is completely determined by squark profile, it is useful to interpret one factor of $\rho$ as  part of the expansion parameter $gq^*\rho$, while the second factor of $\rho$ is the consequence of the requirement that the zero mode is normalized to $1$.

After combining all the factors and performing the integral over Grassmannian variables one finds for two point correlation function
\beq
 \langle\chi (x_1) \bar \chi (x_2) \rangle\propto\int d^4x_0 d\rho\rho^{4N-1}\frac{\mu_0^{2N+1}\left( q^*\overline{q}^*\right)^{N}}{g^{2N}}S_4(x_1-x_0) \, S_4 (x_2 - x_0) \exp\left[-\frac{8\pi^2}{g^2}-4 \pi^2\rho^2|q|^2\right]~. \nonumber \\
\eeq
Integrating over the instanton size gives the effective fermion mass
\beq
 m_\chi\propto \frac{\mu_0^{2N+1}}{\left(gq\right)^{2N}} e^{-8\pi^2/g^2}~,
\eeq
where $g^2$ should be interpreted as the coupling renormalized at the cutoff $\mu_0$. Rewriting this result in terms of the RG invariant scale 
\beq
 \Lambda^{2N+1}=\frac{\mu^{2N+1}}{g^{2N}(\mu)}\exp\left[-8\pi^2/g^2(\mu)\right]
\eeq
and recalling that the holomorphic mass term is given by the second derivative of the superpotential one finds
\beq
 W_{\rm ADS}=\frac{\Lambda^{2N+1}}{\det{Q{\overline Q}}}~.
\eeq

\subsection{Multi-monopole contributions to the path integral}

We are now ready to apply path integral techniques  to the calculation of the correlation functions in multi-monopole backgrounds on ${\bf R}^3\times {\bf S}^1$. To evaluate this contribution, we will consider a specific example, an $SU(3)$  SUSY gauge theory with one flavor, and go through the steps corresponding to those of the instanton calculation reviewed above. The case of $SU(N)$ with one flavor would be completely analogous. We will comment on the case with general number of flavors at the end of this section. We will consider the region (\ref{coulombsubbranch}) of the moduli space, where the two $U(1)$'s are broken to the diagonal subgroup, giving rise to confinement of the two fundamental monopoles. While monopole confinement is an intrinsically non-perturbative effect, the semiclassical approximation presented here will provide additional support for this effect. In addition to confining the monopoles, the squark VEVs will lift some of the fermionic zero modes and partly rotate the remaining exact zero modes into the quark fields, allowing the generation of a holomorphic superpotential term. In our case the relevant multi-monopole configuration is a background containing two distinct fundamental monopoles (of size $1/(g v)$) separated by a distance $\rho \gg 1/(g v)$. Thus the effective size of the multi-monopole is $\rho$. Just as in the 4D instanton case, we will  keep the multi-monopole size fixed, integrating over it at the end of the calculation. Imposing the constraint on the multi-monopole size allows us to turn on an asymptotic squark VEV. The magnetic flux corresponding to the broken $U(1)_X$ is confined in a flux tube with a width that is set by the inverse of the mass of the broken $U(1)_X$ gauge boson $\sim 1/(g|q|)$, since by D-flatness $|{\overline q}| = |q|$. 
As we will see the path integral is dominated by monopole separations of order
\beq
\rho \sim \frac{1}{R|q|^2}~,
\eeq
so in the regime of weak coupling, $g \ll R|q|$, we should consider the case when the  distance between the monopoles, $\rho$, is much smaller than the flux tube width  $1/(g|q|)$. For small enough values of $|q|$,  $\rho$ can still be large compared to the size of the individual monopoles, $1/(gv)$. Then, in the region where there is a significant magnitude for the broken gauge field due to the two monopoles, the broken gauge field itself is approximated by a 3D dipole. If the dipole is oriented along the $z$ axis and centered around the point ${\vec x}_0$ then, since the monopoles have opposite charges under $U(1)_X$, we have
\beq
A_{Xi}(\vec{x}) \propto \frac{{\hat r}_1}{g\, |\vec{x}-\vec{x}_0-\frac{1}{2}\rho{\hat z}|}-\frac{{\hat r}_2}{g\, |\vec{x}-\vec{x}_0+\frac{1}{2}\rho{\hat z}|}~.
\label{dipole}
\eeq
where ${\hat r}_1$ and ${\hat r}_2$ are the unit vectors pointing pointing towards $\vec{x}$ from the position of the corresponding monopole.
The field is concentrated in the region (of size $\rho$) between the two monopoles where
\beq
A_{Xi}  \propto \frac{2\,{\hat z}}{g\, |{\vec x}-\vec{x}_0|}~,
\label{dipolemiddle}
\eeq
while at large distances it falls much more quickly:
\beq
|A_{Xi} | \propto \frac{\rho}{g\, |{\vec x}-\vec{x}_0|^2}~.
\label{dipolefar}
\eeq
This gauge field is independent of the compactified direction, and we will assume that we are in a region of the mixed Coulomb branch where the squark VEVs do not lift the adjoint scalar, $A^0$,  as in  Eq. (\ref{coulombsubbranch}).
Then the contribution of the mass term in the Lagrangian to the classical action is (also integrating over the compact direction)
\beq
\int d^4x\, g^2 A_X^\mu({\vec x}) A_{X\mu}({\vec x})  Q^\dagger({\vec x})  Q({\vec x})  &\propto & R \int_{|x|<\rho} d^3x\, \left(\frac{1}{g\, |{\vec x}|}\right)^2 g^2 |q|^2 \propto R \rho |q|^2\ .
\eeq

This gives us the classical action of the two-monopole configuration in the presence of squark VEVs:
\beq
\label{eq:twomonopoleS}
 S(\rho)=\int d^4x |F_{\mu\nu}|^2+  g^2 |A_\mu q|^2= \frac{8\pi^2 Rv}{g^2}+a |q|^2\rho R\,,
\eeq
where $a$ is a numerical factor. The linear dependence of this action on $\rho$ indicates that the monopoles are confined. However, in an analogy with the constrained instanton calculation of \cite{Affleck:1983mk}, we will allow for an arbitrary inter-monopole distance $\rho$ and integrate over $\rho$ at the end of the calculation.

The two-monopole configuration has $8$ bosonic zero modes. Of these, four zero modes are collective coordinates that correspond to the location of the center of the two-monopole configuration, $x_i$, and its size $\rho$. The bosonic zero mode measure in the path integral  is then
\beq
 d\mu_B\propto \frac{d^3x_0 d\rho}{\rho^4}\left(\frac{(\rho R)^{1/2}\mu_0}{g}\right)^{8}~.
\eeq
Just as in the 4D instanton case, each bosonic zero modes contributes a factor of $\mu_0/g$. A new factor of $\sqrt{R}$ enters through the normalization of the zero modes and is a consequence of the fact that fields are independent of the compact dimension.

The contribution of the lifted fermionic zero modes (corresponding to the closed fermion lines in Fig.~\ref{fig:multiinst}) is identical to the 4D case (\ref{fermionzeromodemeasure}):
\beq
 d\mu_F\propto d\xi d\bar\xi\left(\frac{g^2 q^*\, {\overline q}^*}{\mu_0^2}\right)\frac{1}{\mu_0}\,.
 \label{3dfermionzeromodemeasure}
\eeq

Finally  we need to discuss the structure of two exact zero modes. These are mostly supersymmetric gaugino zero modes living in the unbroken $U(1)_Q$. However, as explained in Section \ref{sec:SU3} and Appendix \ref{app:Effective}, for $\rho\ne 0$ the supersymmetric gaugino zero modes have a small admixture of $U(1)_X$ gauginos. The $U(1)_X$ component of gaugino zero modes is short range (falling off as $1/x^3$), see (\ref{eq:QXbasis}) and at large distances from  the the monopoles, the zero modes behave as the zero modes of a single composite monopole associated with $U(1)_Q$
\begin{equation}
 \lambda^{(0)}_Q(x-x_0)\propto \frac{\rho^{1/2}}{R^{1/2}(x-x_0)^2}\,\xi\propto \sqrt{\frac{\rho}{R}}\,S_3(x-x_0)\,\xi\,,
 \label{3Dzero}
\end{equation}
where the factor $\sqrt{\rho/R}$ arises due to the requirement that the zero mode is normalized to $1$. 

Despite their short range behavior, the $U(1)_X$ components of the supersymmetric gaugino zero modes have an important consequence: these anti-chiral modes mix with chiral components of matter fermions. Indeed, just as in 4D the anti-chiral matter fields satisfy (\ref{antichiraleom}). The right hand side of (\ref{antichiraleom}) is non-vanishing precisely due to the $U(1)_X$ component of the gaugino zero modes. Moreover, the $1/x^3$ behavior of this term implies that the quark and anti-quark components of the zero modes are long range
\beq
 \chi^{(0)}_X(x)\propto\frac{gq^*\rho^{3/2}}{R^{1/2}(x-x_0)^2}\,\xi\propto \frac{gq^*\rho^{3/2}}{R^{1/2}}\,S_3(x-x_0)\,\xi~,
 \label{Xzeromode}
\eeq
where, just as in the 4D case the factor $gq^*\rho$ is the expansion parameter and the additional factor of $\rho^{1/2}$ arises from the normalization of the zero mode.

We can now combine all the factors to calculate the gaugino-gaugino and quark-quark correlation functions. For quarks we project the zero mode legs onto the massless matter fermion in (\ref{Xzeromode}) and after integration over Grassmannian variables we find
\beq
 \langle \chi_X(x){\overline \chi}_X(y)\rangle \propto e^{- 4\pi^2 v R/g^2}\left(\frac{R^3\mu_0^5 }{g^4	}\right)\left(q^* {\overline q}^*\right)^2\int d^3x_0 d\rho\rho^3   \,S_3(x-x_0) S_3(y-x_0) \,e^{-a R |q|^2\rho}~.
 \label{eq:correlationfunction}
\eeq
Integrating over the multi-monopole size $\rho$ we find
that the path integral is dominated by 
\beq
\rho \sim \frac{1}{ R|q ||{\overline q}|}\,.
\eeq 
 Finally, setting the cutoff of the 3D theory at $\mu_0=1/R$ we obtain
\beq
 \langle \chi_X(x){\overline \chi}_X(y)\rangle\propto e^{- 8\pi^2R v/g^2} \int d^3x_0 \frac{S_3(x-x_0)S_3(y-x_0)}{g^4 R^{6}q^2\overline q^2}  ~.
\eeq
This corresponds to a fermion mass term that is a holomorphic function of $q$ and ${\overline q}$:
\beq
 m_\chi\propto\left(\frac{1}{g^4 R^5}\right) \frac{e^{- 8\pi^2Rv/g^2}}{q^2 \,{\overline q}^2}~.
\eeq
Since this mass term is holomorphic it should arise as the second derivative of the superpotential 
\beq
m_\chi =\frac{\partial^2 W}{\partial Q \,\partial{\overline Q}}~,
\eeq
 from which we can identify the two-monopole contribution to the 3D superpotential
\beq
\label{eq:multimonopole}
 W_{\rm pre-ADS}\propto\frac{e^{- 8\pi^2Rv/g^2}}{R^4 Q \,{\overline Q}}\propto \frac{1}{Y_1Y_2 Q{\overline Q}}\,,
\eeq
in agreement with (\ref{eq:effectiveW}).

For the gaugino-gaugino correlation function we project onto the zero mode component (\ref{3Dzero}) to find
\begin{equation}
\begin{split}
 \langle \lambda(x)\lambda(y)\rangle &\propto e^{- 8\pi^2 v R/g^2}\int d^3x_0 d\rho \left(\frac{R^4\mu_0^5 q^*\overline q^*}{g^6} \right)  \,\lambda_Q^{(0)}(x) \lambda_Q^{(0)}(y) e^{-a R |q|^2\rho}\\
&\propto e^{- 8\pi^2 v R/g^2}\int d^3x_0 d\rho\rho \left(\frac{R^3\mu_0^5 q^*\overline q^*}{g^6} \right)  \,S_3(x-x_0)S_3(y-y_0) e^{-a R |q|^2\rho}\\
  &\propto e^{- 8\pi^2R v/g^2} \int d^3x_0 \frac{S_3(x-x_0) S_3(y-x_0)}{g^6 R^4 q{\overline q}} ~.
 \label{eq:QQcorrelationfunction}
\end{split}
\end{equation}
The  corresponding gaugino mass term is
\beq
 m_\lambda \propto\frac{1}{g^6 R^3} \frac{e^{-8\pi^2Rv/g^2}}{q \,{\overline q}}\equiv \frac{\partial^2 W}{\partial \phi \,\partial \phi}~,
\eeq
We conclude that both quark-quark and gaugino-gaugino correlation functions imply the same superpotential (\ref{eq:multimonopole}).

The calculations of this section can be generalized to $(F+1)$-monopole configurations in theories with $F$ flavors. The only non-trivial step in such a generalization is the introduction of the collective coordinates describing the multi-monopole configuration, one combination of which will correspond to the size. While the size is easily seen to be the  inter-monopole distance in the two-monopole case, the relevant definition in the multi-monopole case is more complicated.


\section{Confined Monopoles in $SU(N)$}

\label{sec:SUN}
In this section we generalize the discussion of the pre-ADS superpotential generated by multi-monopoles  to the general case of $SU(N)$ with $F$ flavors. In the absence of squark VEVs we can take a generic VEV for $\phi$
\beq
\phi={\rm diag}(v_1,\ldots,v_N)
\eeq
 and the gauge symmetry is broken to $\Pi_{i=1}^{N-1} U(1)_i$. Choosing for concreteness a region of the fundamental  Weyl chamber satisfying $v_1>0>v_2>\ldots>v_N$, one finds that single monopole contributions to the superpotential are

\beq
W= \eta \Pi_i Y_i +\sum_{i=2}^{N-1} \frac{1}{Y_i}\,.
\eeq
Since quark zero modes are localized on the first fundamental monopole, its contribution is missing from the superpotential.
Next  we turn on generic squark VEVs. To be able to do so we must set $F$ diagonal elements of the adjoint VEV to zero---indeed, single monopole effects push the theory precisely in this direction. As a result the squark VEVs can appear (by a gauge choice) in colors 2 up to $F+1$, and 
the gauge symmetry is broken to $\Pi_{i=F+2}^{N-1} U(1)_i$.  The squark VEVs confine the first $F+1$ fundamental monopoles into a composite which naively inherits $2(F+1)$ gaugino and $2F$ quark zero modes of its constituents. The Coulomb branch will now be described by the modulus corresponding to the confined multi-monopole given by $\Pi_{i=1}^{F+1} Y_i$ as well as the remaining $Y_i$ moduli (where $i>F+1$) that are not confined. The resulting table of symmetry charges for this general case with maximal rank squark VEVs is given by
\beq
\begin{array}{|c|cccccc|} \hline
& SU(N) & SU(F) & SU(F) & U(1)_A & U(1)_B & U(1)_R \\ \hline
Q & \Yfund & \Yfund & 1 & 1 & 1 & 0 \\
\bar{Q} & \overline{\Yfund} & 1 & \Yfund & 1 & -1 & 0 \\ 
\Pi_{i=1}^{F+1} Y_i & 1 & 1 & 1 & -2 F & 0 & -2 \\
Y_{F+2} & 1 & 1 & 1 & 0 & 0 & -2 \\
\vdots & \vdots & \vdots & \vdots & \vdots & \vdots & \vdots \\
Y_{N-1} & 1 & 1 & 1 & 0 & 0 & -2 \\ \hline 
\end{array}
\eeq
The most general  superpotential allowed by these symmetries is
\beq
W_{\rm pre-ADS}= \eta \Pi_i Y_i + \frac{1}{{\rm det}\,Q \overline{Q}\,\Pi_{i=1}^{F+1}Y_i } +\sum_{i=F+2}^{N-1} \frac{1}{Y_i}~.
\label{eq:preADS2}
\eeq
The middle term is again to be interpreted as the contribution of the $F+1$ confined monopoles\footnote{For the case of $F=N-2$, the semiclassical field configuration of the composite monopole is actually known explicitly   \cite{Weinberg:1979zt,Weinberg:1998hn}.}. It replaces the $1/Y_i$ contributions of $F+1$ single fundamental monopoles whose individual actions become infinite in the presence of squark VEVs. In terms of zero mode counting we can interpret the multi-monopole term in the following way: all but two zero modes of the confined multi-monopole can be closed off with with the insertion of $F$ squark and $F$ antisquark VEVs to obtain (\ref{eq:preADS2}). The first term is the effect of the KK monopole, which breaks some of the global symmetries (the ones that are anomalous in the 4D theory). The final sum consists of the usual AHW single monopole superpotential terms induced on the Coulomb branch.  

Integrating out the lifted Coulomb moduli $Y_{F+2}, \ldots , Y_{N-1}$ we obtain the globally defined 3D superpotential of~\cite{Aharony:1997bx}:
\beq
\label{eq:3DWADS}
 W_{\rm 3D}= \eta Y + (N-F-1) \frac{1}{(Y\,{\rm det}Q {\overline Q})^{\frac{1}{N-F-1}}}\ .
\eeq
Of course, this superpotential also lifts the global moduli $Y$ and $Q\bar Q$. Integrating out the monopole modulus first, we find the ADS superpotential of the theory on $\mathbf{R}^3\times \mathbf{S}^1$:
\beq
\label{eq:WADS}
W_{\rm ADS} =(N-F)\left(\frac{\Lambda^{3N-F}}{\det Q\overline{ Q}}\right)^\frac{1}{N-F}\,,
\eeq
where we replaced $\eta$ with $\Lambda^{3N-F}$ in anticipation of taking the infinite radius limit. Indeed at this point such a limit is trivial.
Thus we see that for the general case the origin of the ADS superpotential can be traced back to an $F+1$ multi-monopole contribution to the superpotential in the theory on a circle. We expect that there should be an equivalent field configuration contributing to the path integral for fully 4D theories as well. It would be very interesting to explore the exact nature of that multi-monopole for the full theory.

Now let us consider the case with large squark VEVS: $q,{\overline q}> 1/R$. In this case the gauge group is broken in the 4D regime so we first match the $SU(N)$ gauge theory with $F$ flavors onto the low-energy effective theory with $SU(N-F)$ and no flavors.
The matching in terms of the strong coupling scale is
\beq
\Lambda^b=\Lambda^{3N-F} = \Lambda_L^{3(N-F)} (Q{\overline Q})^{F}=\Lambda_L^{b_L} \,{\rm det}\,Q{\overline Q}~.
\label{scalematchingN,F}
\eeq

On ${\bf R}^3\times {\bf S}^1$ the $SU(N-F)$ gauge theory with no flavors has $N-F-1$ fundamental monopoles, so the superpotential is
\beq\label{Mario1}
W=\eta_L Y_L +\sum_{j=F+1}^{N-1}\frac{1}{Y_{L,j}}~.
\eeq
Note that \eqref{scalematchingN,F} implies $\eta_L=\Lambda^{b_L}_L=\Lambda^b/{\rm det}Q\bar{Q}$. Then integrating out all the lifted $Y_{L,j}$ Coulomb branch directions we can easily check that \eqref{Mario1} reproduces the ADS superpotential as expected.

Furthermore it is convenient to relabel the summation index by $i= j+F$, so that the sum runs from  $F+1$ to $N-1$.
Then semiclassically (in terms of the relabelled indices and $SU(N)$ VEVs and the roots $\alpha_i$)
\beq
Y_{L,i}&=&e^{8\pi^2 vR \,({\bf h}\cdot {\bf \alpha_i})/g_{L}^2}~, \quad i>F+1 \\
Y_{L,F+1}&=&e^{8\pi^2 vR \,({\bf h}\cdot \sum_{i=1}^{F+1}{\bf \alpha_i})R/g_{L}^2}~, \\
\eta_L Y_L&=&e^{-8\pi^2/g_{L}^2}e^{8\pi^2 vR\, ({\bf h}\cdot \sum_{i=1}^{N-1}{\bf \alpha_i})/g_{L}^2}=e^{-8\pi^2/g_{L}^2}e^{8\pi^2 vR {\bf h}\cdot {\bf \alpha_0}/g_{L}^2}~.
\eeq
so we see that  identifying 
\beq
Y_L={\rm det}\,Q{\overline Q}\Pi_i Y_i~, \quad Y_{L,F+1}={\rm det}\,Q{\overline Q} \Pi_{i=1}^{F+1} Y_i ~,\quad {\rm and} \quad \eta_L  =\eta/{\rm det}\,Q{\overline Q}~,\eeq 
gives the two superpotentials have the same dependence on the squark VEVs.

Finally let us consider the case of $SU(N)$ with $F=N-1$ flavors. In this case turning on the maximal rank squark VEVs breaks all the $U(1)$'s. Thus the Coulomb branch is completely lifted and low energy degrees of freedom do not contain the monopole modulus $Y$. Despite the absence of a true monopole, there still exists a topologically non-trivial field configuration which can be interpreted as an $N$-monopole configuration containing each fundamental monopole as well as the KK monopole. Indeed, this $N$-monopole configuration is a periodic instanton \cite{Harrington:1978ve} of the theory on a circle and it turns into a conventional instanton in the infinite radius limit. This $N$-monopole configuration contributes to the path integral and the superpotential. Its contribution could be determined by using the symmetry arguments of this section or calculating multi-monopole contributions to the path integral as in Sec. \ref{sec:pathintegral}. In either case one finds
\beq
\label{eq:inst}
W =  \frac{\eta \Pi_i Y_i}{\Pi_i Y_i\,\det Q \overline{Q} } =   \frac{\eta }{\det Q \overline{Q}} =\frac{\Lambda^{3N-F}}{\det Q\overline Q}\,,
\eeq
which is exactly the expected 4D instanton induced ADS superpotential term for $F=N-1$.


\section{Conclusions}
\label{sec:conclude}
Examples of dynamically generated superpotential terms with a clear underlying dynamical origin are few and far between. 4D instantons only contribute in special cases, while the other known examples are usually obtained using indirect methods. In this paper we have established that the confinement of monopoles is the underlying dynamical origin for the generic ADS superpotential when the 4D theory is compactified on a circle.  This happens on the mixed Higgs-Coulomb branch, where turning on some of the squark VEVs breaks one or more $U(1)$ gauge groups, leading to the confinement of some of the fundamental monopoles. At the same time most of the fermionic zero modes of the multi-monopole are lifted. In particular, we have identified an $F+1$ multi-monopole for the case of $SU(N)$ SUSY QCD (with four supercharges) and $F$ flavors. In the presence of squark VEVs these monopoles are confined, and exactly two fermionic zero modes remain unlifted (these modes being partly in the quarks and partly in the gauginos). The resulting pre-ADS superpotential term is inversely proportional to the fields but no fractional powers appear. Since it is a contribution on the mixed Higgs-Coulomb branch it depends both on the Coulomb and the Higgs moduli. The globally defined 3D ADS superpotential can be obtained by integrating out the lifted Coulomb moduli $Y_i$, while the full 4D ADS superpotential is obtained by also integrating out the global Coulomb modulus $Y=\prod_i Y_i$. A symmetry argument clearly shows that the pre-ADS superpotential can be generated. We have presented  a detailed accounting of the path integral calculation of the fermionic two-point functions in the presence of the confined multi-monopole, argued that these indeed correspond to the presence of the pre-ADS superpotential, and that they yield a more detailed dynamical explanation of the origin of the well-known ADS superpotential.

\section*{Acknowledgements}

The authors thank Ofer Aharony and Ken Intriligator for useful discussions. We thank the Aspen Center for Physics - supported in part by NSF PHY-1607611 - and the  Mainz Institute for Theoretical Physics (MITP) for hospitality and partial support during the completion of this work.
C.C. is supported in part by the NSF grant PHY-1719877. Y.S. is supported in part by NSF grant PHY-1620638. J.T is supported in part by DOE grant DE-SC0009999.


\appendix 


\section*{Appendix}



\section{Monopole solutions for $SU(3)$\label{app:monopolesolution}}

Above we have presented our detailed dynamical explanation for the generation of the ADS superpotential \cite{Affleck:1983mk} in $SU(3)$ theories due to multi-monopole dynamics. To remind the reader of the background we will review the details of $SU(3)$ monopoles.
The 't Hooft-Polyakov monopole solution \cite{`tHooftPolyakov} gives a configuration with a magnetic charge in an $SU(2)$ gauge theory that is broken to $U(1)$ by an adjoint VEV. In the absence of a scalar potential (a.k.a. the BPS limit) the fields (in the hedgehog gauge) are given by \cite{PrasadSommerfield}
\beq
\phi^a = \hat{r}^a v f(r,v)~,\quad
A^a_i=\epsilon^{ai j} \,\hat{r}^{\,j} \frac{k(r,v)}{g\,r}
\eeq
where $f(r,v)$ and  $k(r,v)$ approach 1 as $r\rightarrow \infty$. The solutions are:
\beq
f(r,v)&=& \coth(g v r)- \frac{1}{g v r}~,\\
k(r,v) &=& 1-\frac{g v r}{\sinh(g v r)}~.
\label{monopolesolutionprofile}
\eeq

Using this we can also easily find the monopole solutions in an $SU(3)$ gauge group \cite{Weinberg:1979zt,Weinberg:1998hn}. By a gauge choice we will work in the fundamental Weyl chamber
\beq
\phi={\rm diag}(v_1,v_2,v_3)
\label{eq:VEV}
\eeq
where $v_1 > v_2>v_3$ and $v_1 + v_2+v_3=0$.
On the Coulomb branch with the squark VEVs set to zero $\langle Q \rangle=\langle \overline{Q}\rangle=0$ the gauge symmetry is broken to $U(1)_1\times U(1)_2$. The unbroken $U(1)$ charge generators are:
\beq
Q_1&=&\frac{1}{2}\,{\rm diag}(1,-1,0)  \\
Q_2&=&\frac{1}{2}\,{\rm diag}(0,1,-1) ,
\label{eq:unbroken}
\eeq
 
To find the monopole solutions we can split the VEV (\ref{eq:VEV}) into two pieces that correspond to an adjoint and a singlet  under the $SU(2)$ subgroup in question.  For example for the $SU(2)$ subgroup corresponding to $Q_1$ we can write
\beq
\phi&=&{\rm diag}\left(\frac{v_1-v_2}{2},-\frac{v_1-v_2}{2},0 \right) +
{\rm diag}\left(\frac{v_1+v_2}{2},\frac{v_1+v_2}{2},-v_1-v_2 \right) \equiv \tilde{v}_1 + w_1
\label{eq:VEVsplit}
\eeq
and similarly for $Q_2$ yielding the corresponding $\tilde{v}_2, w_2$ diagonal matrices. Then we can write the two BPS monopole solutions as \cite{Weinberg:1979zt,Weinberg:1998hn}:
\beq
 \phi_i(r)= \hat r^a \tau_i^a v  f(r,v) +w_i
\label{eq:embed}
\eeq
where, $v=2 \sqrt{v_1^2 + v_1 v_2 + v_2^2}$, and $\tau_i^a$ are generators of the $SU(2)$ with diagonal generator $Q_i=\tau^3_i$.

\section{Roots of $SU(N)$\label{app:roots}}

In order to generalize our results to $SU(N)$ let us review how the results previously obtained can be written in terms of simple roots. First let us start from $SU(3)$.
 On the Coulomb branch with the squark VEVs set to zero $\langle Q \rangle=\langle \overline{Q}\rangle=0$ the gauge symmetry is broken to $U(1)_1\times U(1)_2$, the unbroken $U(1)$ charge generators can be chosen to correspond to the two simple roots, ${\bf \alpha_i}$, which are, explicitly 
 \beq
 {\bf \alpha_1}=(1,0),\quad\quad {\bf \alpha_2}=(-1/2,\sqrt{3}/2)~,
 \eeq
 and the two Cartan generators, $H_i$, of $SU(3)$:
 \beq
H_1={\rm diag}\left( \frac{1}{2}, -\frac{1}{2},0\right), \quad\quad H_2={\rm diag}\left( \frac{1}{2\sqrt{3}}, \frac{1}{2\sqrt{3}},-\frac{1}{\sqrt{3}}\right)~.
\eeq
\begin{figure}[h!]
\centering
\includegraphics[height=4cm]{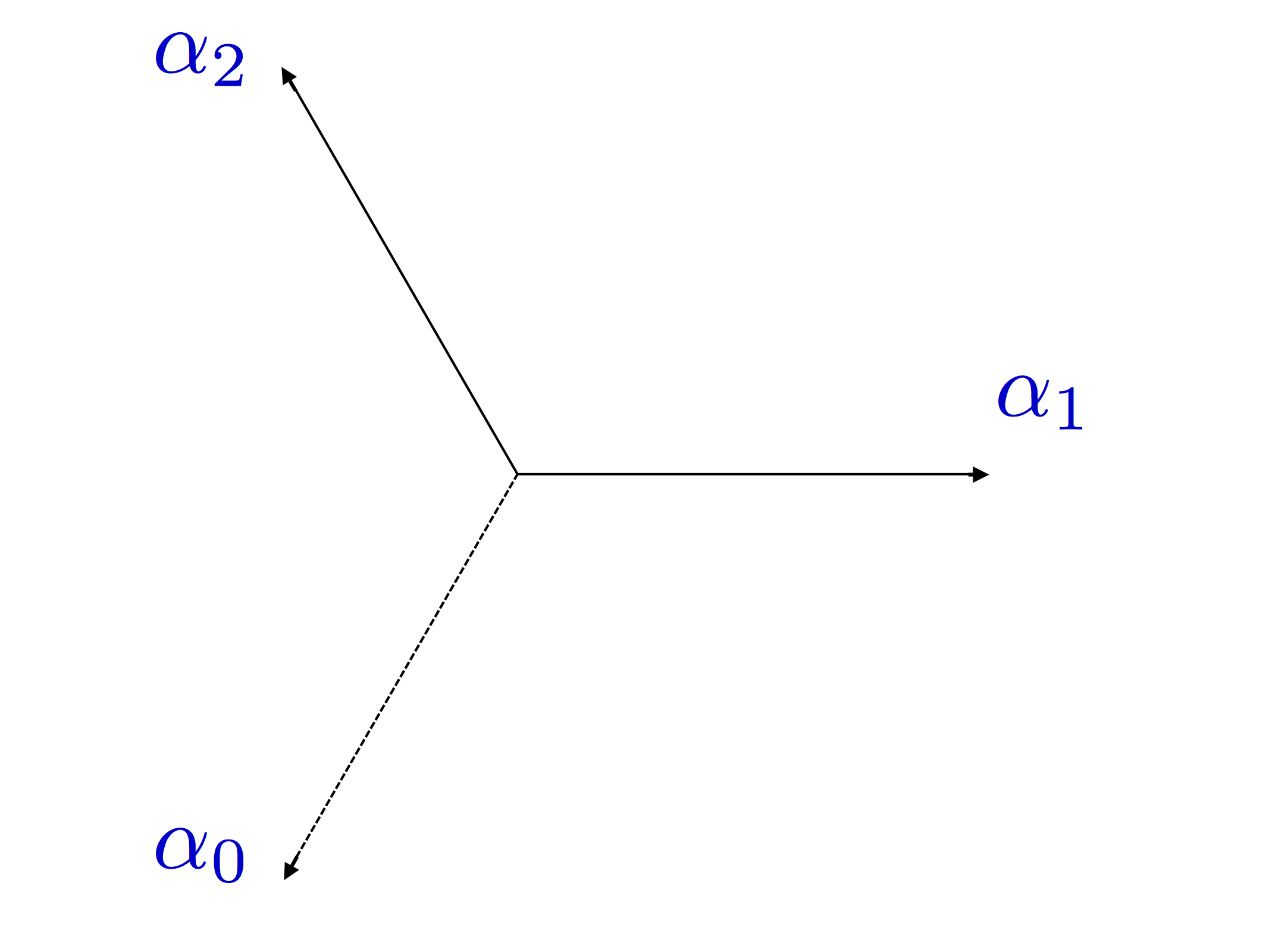}\caption{Simple roots, $\alpha_1$ and $\alpha_2$, of $SU(3)$, and the lowest negative root $\alpha_0$.}
\label{fig:roots}
\end{figure}

Each element of an $SU(3)$ representation can be assigned a charge under the two $U(1)$'s.  For example, the fundamental representation has
charges
\beq
\left(\begin{array}{c} r \\g \\ b \end{array} \right)\sim \left(\begin{array}{c} \left(\frac{1}{2}, \frac{1}{2\sqrt{3}}\right)\\ \left(-\frac{1}{2}, \frac{1}{2\sqrt{3}}\right) \\ \left(0, -\frac{1}{\sqrt{3}}\right)\end{array}\right)~.
\eeq
Then we see that starting with the lowest state,
\beq
b+{\bf \alpha_2}\sim g~,\quad\quad g+{\bf \alpha_1}\sim r~,
\eeq
we can work our way up through the entire representation using the roots.
Thus the roots represent the charges of the off-diagonal generators (the analogs of the $W$'s) that can make the transition from one element of a represent to another element (aka ladder operators).

We can assemble the $H_i$ generators into a vector ${\bf H}$ and then write the unbroken $U(1)$ charge generators in the conventional basis associated with the two BPS monopoles as
\beq
Q_1&=&\frac{1}{2}\,{\rm diag}(1,-1,0) ={\bf \alpha_1}\cdot{\bf H}~,\nonumber \\
Q_2&=&\frac{1}{2}\,{\rm diag}(0,1,-1)={\bf \alpha_2}\cdot{\bf H}~,
\label{eq:unbrokenCart}
\eeq
Expressing the unbroken generators (\ref{eq:unbrokenCart}) in terms of the roots, allows us to immediately read off how a given 't Hooft-Polyakov monopole solution, which is associated with a breaking of a particular $SU(2)$ factor, is embedded into the full $SU(N)$ gauge group  \cite{Weinberg:1979zt,Weinberg:1998hn}. 

Consider $Q_1=\alpha_1\cdot{\bf H}$, we can rewrite Eq. (\ref{eq:VEVsplit}) in terms of the Cartan elements as:
\beq
\phi =  v\, {\bf h}\cdot {\bf H}
\eeq
where
\beq
h=\left(\frac{v_1-v_2}{v},\frac{\sqrt{3} (v_1+v_2)}{v}\right)~,\quad\quad v=2 \sqrt{v_1^2 + v_1 v_2 + v_2^2}~.
\eeq
Note that ${\bf h}\cdot {\bf h}=1$. The condition that we are in the fundamental Weyl chamber ($v_1 > v_2>v_3$) is
\beq
{\bf h} \cdot \alpha_i \ge 0~.
\label{eq:weylchamber}
\eeq
With this notation we see that the first term in (\ref{eq:VEVsplit}) is just
\beq
{\tilde v}_1=v\,\left( {\bf h}\cdot \alpha_1\right)\left( \alpha_1 \cdot {\bf H}\right)={\rm diag}\left(\frac{v_1-v_2}{2},-\frac{v_1-v_2}{2},0 \right) ~.
\eeq
The second term in  (\ref{eq:VEVsplit}) is just the remainder:
\beq
\phi -{\tilde v}_1=v( h -  (h \cdot \alpha_1) \alpha_1 )\cdot {\bf H})~.
\eeq
Then we can write the BPS monopole solutions as
\beq
 \Phi_i= \hat r^a \tau^a v (h \cdot {\bf \alpha_i}) f(r,v (h \cdot {\bf \alpha_i})) +v( h -  (h \cdot {\bf \alpha_i}) {\bf \alpha_i} )\cdot {\bf H})
\label{eq:embedgen}
\eeq
where, $\tau_i^a$ are generators of the $SU(2)$ associated with ${\bf \alpha_i}$.

This embedding pattern can be repeated for any $SU(N)$. For each simple root ${\bf \alpha_j}$ of the $SU(N)$ gauge group there is an $SU(2)$ subgroup whose diagonal generator is ${\bf \alpha_j} \cdot {\bf H}$.
With a general gauge group the BPS monopoles corresponding to the root ${\bf \alpha_i}$  have a magnetic charge vector\footnote{The $j$th component of the charge vector gives the charge associated with $Q_j={\bf \alpha_j} \cdot {\bf H}.$} \cite{Weinberg:1979zt,Weinberg:1998hn}  ${\bf \alpha_i}^*={\bf \alpha_i}/{\bf \alpha_i} \cdot {\bf \alpha_i}$. For $SU(N)$ this simplifies since ${\bf \alpha_i}^*= {\bf \alpha_i}$. The KK monopole is associated with the lowest negative root, $\alpha_0$, and has magnetic charge 
$\alpha_0^*=\alpha_0=-\sum_j{\bf \alpha_j}$.


 \section{The Coulomb branch in 3D ${\cal N}=2$ SUSY\label{app:3DSUSY}}
 
$\mathcal{N}=2$ SUSY QCD in 3D can be obtained by a dimensional reduction of the corresponding $\mathcal{N}=1$ 4D theory (both are theories with four supercharges). After Wick rotation, compactification of the time direction, and Wick rotation of a spatial direction, the 4D time component of the gauge field turns into a scalar in the adjoint representation, $A_0\rightarrow\phi$, and as a result the 3D theory acquires a Coulomb branch. On the Coulomb branch, the photons of the unbroken $U(1)$ gauge symmetries can be dualized to a scalar: $\epsilon_{ijk}F^{jk}\sim \partial_i \gamma$. The effective low energy Lagrangian on the Coulomb branch can be written in terms of the chiral superfield $\varphi=\phi+i\gamma$ and its superpartners, where $\phi=A_0$. There are instantons on the Coulomb branch of non-Abelian gauge theories in 3D and their field configurations can be obtained from a dimensional reduction of a 4D monopole with its worldline wrapped around the compactified dimension (thus we usually refer to 3D instantons as monopoles). The fundamental result on which both the current paper and much of the past work is based was obtained by AHW \cite{Affleck:1982as}: they found that in $\mathcal{N}=2$ SUSY pure Yang-Mills in 3D  the gaugino two point correlation function has non-trivial contribution in a wrapped monopole background.  In an $SU(2)$ theory with no matter this results in a superpotential
\beq
 W=\frac{1}{Y}~,
\eeq
where $Y$ is the monopole operator parameterizing the Coulomb branch  
\beq
Y=\exp{\left[8\pi^2(\varphi_{11}-\varphi_{22})R/g^2\right]}~.
\eeq

The above result can be easily generalized to theories with larger gauge symmetry. For example, in $SU(N)$ the classical Coulomb branch is $N-1$ dimensional. At the generic point on the classical Coulomb branch the $SU(N)$ is broken to $U(1)^{N-1}$ and the field $\varphi$ acquires a VEV $\langle\varphi_{ii}\rangle=v_i$. It is convenient to choose the basis in which generators of these  $U(1)$'s are given by $T_j=1/2\,\, \mathrm{diag}(0,\ldots,1,-1,\ldots,0)=\boldsymbol{\alpha}_j\cdot {\bf H}$, where $1$ appears in the $j$th slot along the diagonal.  Each ordering of VEVs defines a so called Weyl chamber and it is sufficient to consider dynamics in the fundamental Weyl chamber which is defined by 
\beq
v_1>v_2>\ldots>v_N~.
\label{eq:weylchamber2}
\eeq

One can construct a fundamental monopole associated with each of the $N-1$ $U(1)$'s by embedding them into $N-1$ linearly independent but non-orthogonal $SU(2)$ subgroups. This can be easily done in terms of the simple roots of the initial gauge group (see Appendix~\ref{app:roots}). We see that $\boldsymbol{\alpha}_j \cdot \boldsymbol{F}^{np}$ is the $U(1)$ field strength associated with the Cartan generator $\boldsymbol{\alpha}_j \cdot \boldsymbol{H}$.
We can also write the general moduli in terms of the roots, if we promote $h$ to be a complex field:
\beq
Y_j=e^{8\pi^2 v(\boldsymbol{h}\cdot\boldsymbol{\alpha_j} )R/g^2}~,
\eeq
with 
\beq
 \phi=v \,{\rm Re}(\boldsymbol{ h})\cdot \boldsymbol{H}~,\quad\quad v\, {\rm Im}(\boldsymbol{ h})\cdot\boldsymbol{\alpha}_j=\gamma_j ~, 
 \eeq
where $\gamma_j$ represents the dual photon \cite{Affleck:1982as} 
\beq
\partial_m\gamma_j=\epsilon_{mnp}\,\boldsymbol{\alpha}_j \cdot \boldsymbol{ F}^{np}~.
\eeq
This makes the fundamental Weyl chamber condition (\ref{eq:weylchamber2}) equivalent to (\ref{eq:weylchamber}).

 To determine the resulting superpotential terms for the fundamental monopoles one needs to count fermion zero modes \cite{Jackiw:1975fn}, which can be done using the appropriate index theorems \cite{Callias:1977kg,Erich}. Under each $SU(2)$, the gaugino decomposes into an $SU(2)$ adjoint, $2(N-2)$ doublets and $(N-2)^2-1$ singlets. Inside the fundamental Weyl chamber the doublets obtain large real masses from the $\varphi$ VEVS and do not have zero modes. Thus each fundamental monopole has exactly two gaugino zero modes and there are $N-1$ contributions to two point correlation function, resulting in a superpotential
\beq
 W=\sum_{i=1}^{N-1}\frac{1}{Y_i}~.
\eeq


\section{${\bf R}^3$ vs. ${\bf R}^3\times {\bf S}^1$ and  KK monopoles\label{app:KK}}

 It is important to distinguish between a truly 3D theory on ${\bf R}^3$ and a theory on ${\bf R}^3\times {\bf S}^1$ which exhibits 3D behavior at low energies. In the compactified theory on a circle of radius $R$, the 3D gauge coupling is given in terms of the 4D holomorphic  coupling by
\beq
\frac{1}{g^2_3}=\frac{2\pi R}{g^2_4(1/R)}=2 \pi R\left[\frac{1}{g^2_4(\mu_0)} +\frac{b}{8\pi^2} \ln\left( \frac{1}{R\mu_0}\right)\right]~.
\eeq
  The monopole configuration exists both in the theory on ${\bf R}^3$ and a theory on ${\bf R}^3\times {\bf S}^1$. However, the latter theory has another property that is important for our discussion. The Coulomb branch is periodic with a period $1/R$, $\phi\rightarrow \phi+1/R$. To see this for $SU(2)$, notice that the gauge boson KK tower has masses given by $n/R$ for integer values of $n$ and when $\phi$ acquires a VEV 
\beq
\langle \phi \rangle ={\rm diag}(v,-v)
\eeq
 the masses are shifted to $n/R+v$. As $\langle \phi \rangle$ approaches $1/R$ a new KK state becomes massless and the $SU(2)$ symmetry is restored, implying that the Coulomb branch is periodic with a period $1/R$.  

There is another important distinction between the theory on ${\bf R}^3$ and a theory on ${\bf R}^3\times {\bf S}^1$. The latter has an additional monopole configuration, called the KK monopole \cite{Lee:1997vp}, specific to the existence of the additional $S^1$. The KK monopole can be obtained by twisting  the fundamental monopole configuration around the circle with a large anti-periodic gauge transformation $U=\exp(-i x_0 \,\sigma^3/2R)$. While this large gauge transform $U$ is anti-periodic, the resulting gauge-transformed field configurations are periodic since the vector multiplet transforms in the adjoint representation of $SU(3)$. The action of the KK monopole is given by 
$ S_{KK}=(4\pi/R+\phi_{11}-\phi_{22})/g_3^2$. There are two gaugino zero modes in the KK-monopole background as can be seen directly by performing the large gauge transformation on the zero modes of the fundamental monopole \cite{Csaki:2017cqm}. Thus the chiral two-point gaugino correlation function receives a non-trivial contribution in the KK monopole background, resulting in a new superpotential term of the pure $SU(2)$  super-Yang-Mills:
\beq
 W_{KK}=\exp\left(-\frac{4\pi}{g_3^2 R}-\frac{4\pi^2 (\phi_{22}-\phi_{11})R}{g_3^2}\right)=\eta Y~,
\eeq
where we defined $\eta=\exp\left(-4\pi/g_3^2R\right)=\exp\left(-8\pi^2/g^2\right)$. The parameter $\eta$ can be expressed in terms of the dynamical scale of the 4d theory, $\eta\sim\Lambda^b$, where $b$ is a one-loop beta-function coefficient.


\section{Multi-monopole zero modes\label{app:Effective}}
Here we analyze the structure of supersymmetric gaugino zero modes of the $SU(3)$ theory in the two-monopole background. When the squark VEVs are turned off the gaugino zero modes can easily be obtained by performing a supersymmetry transformation on the two-monopole field
\begin{equation}
 \lambda^0(x)\propto \sigma^{\mu\nu}F_{\mu\nu}\, \xi\propto (B_i^a T^a\sigma^{i}+\nabla_i \phi^aT^a\sigma^i)\xi=2B_i^a T^a\sigma^{i}\xi\,,
\end{equation}
where in the last equality we used the fact that  $B_i^aT^a\sigma^i=\nabla_i\phi^a T^a\sigma^i$. Thus, to understand the scaling of the gaugino zero modes we need to study the magnetic field of the two-monopole configuration.

Consider the case of an $SU(3)$ gauge group with two different monopoles centered around the point $\vec{x}_0$, a distance $\rho$ apart.  We can denote the direction of the line between them by the unit vector ${\hat d}$. Then we have two vectors that point from each of the monopoles to an arbitrary point ${\vec r}$:
\beq
{\vec r_1}=\vec{r}-\vec{x}_0-\frac{1}{2}\rho{\hat d}~,\quad\quad {\vec r_2}=\vec{r}-\vec{x}_0+\frac{1}{2}\rho{\hat d}
\eeq
We will also write $r_i=|{\vec r}_i|$.
With this notation the approximate scalar solution \cite{Weinberg:1979zt} for two widely separated monopoles is 
 \beq
\phi^a T^a&=& \sigma\left[ \vec{h}-\vec{\alpha_1} (\vec{h}\cdot  \vec{\alpha_1}) -\vec{\alpha_2} (\vec{h}\cdot  \vec{\alpha_2}) \right]\cdot \vec{H} +
 \hat{r}_1^a T_{(1)}^a (\vec{h}\cdot  \vec{\alpha_1})\,f( r_1,\vec{h}\cdot  \vec{\alpha_1}) 
 \\&&+
 \hat{r}_2^a T_{(2)}^a (\vec{h}\cdot  \vec{\alpha_2})\,f( r_2,\vec{h}\cdot  \vec{\alpha_2}) 
 \\
 &=&\frac{1}{2} \,{\rm diag}(v,0,-v)+ \frac{\vec{r_1}^a}{r_1} T_{(1)}^a  v\, f(r_1, v) + \frac{\vec{r_2}^a}{r_2}  T_{(2)}^a (v)\, f(r_2, v)~.
\label{monopolewide1}
\eeq
At the point ${\vec r}$ the local unbroken $U(1)$'s can be taken to be
\beq
 \frac{\vec{r_1}^a}{r_1} T_{(1)}^a =Q_1=\frac{1}{2}\,{\rm diag}(1,-1,0)~, \quad\quad \frac{\vec{r_2}^a}{r_2}  T^a_{(2)} =Q_2=\frac{1}{2}\,{\rm diag}(0,1,-1)~.
\eeq

Far from the monopoles the asymptotic scalar VEVs (\ref{coulombsubbranch}) split the $SU(3)$ fundamental  into three singlets $(r,g,b)$ and the $SU(3)$ adjoint into two massless singlets  $\lambda_{(1)}^3$  and $\lambda_{(2)}^3$ (corresponding to the unbroken $U(1)$ generators $Q_1$  and $Q_2$) and fields with masses of order $gv$.

The magnetic field of the two-monopole solution is
\beq
B^a_i T^a&=& D_i \phi^a T^a=\sum_{a=1}^3 \frac{\vec{r}^{\,\,i}_1\,\vec{r}^{\,a}_1}{r_1^2} T_{(1)}^a   \,v \, f^\prime(r_1,v)+\sum_{a=1}^3 \frac{ \vec{r}^{\,\,i}_2\,\vec{r}^{\,a}_2}{r_2^2} T_{(2)}^a  \,v \, f^\prime(r_2,v)\nonumber\\
&&+{\mathcal O}\left(\frac{k^\prime(r_1,v)}{g\,|\vec{r_1}|}\right)+{\mathcal O}\left(\frac{k^\prime(r_2,v)}{g\,|\vec{r_2}|}\right)~.
\eeq
From (\ref{monopolesolutionprofile}) we have for large $r$  
\beq
f^\prime(r,v) &\sim& \frac{1}{gv r^2}~,\\
k^\prime(r,v) &\sim& 2 g^2 v^2 \,r \,e^{-gv r}~.
\eeq
For distances $|\vec{r_1}|,|\vec{r_2}| \gg 1/v$ we can neglect the exponentially suppressed terms. At a  given point in space there are two long range magnetic fields whose directions in $SU(3)$ group space are aligned with   $\vec{r_1}^a$ and $\vec{r_2}^a$. For simplicity we can choose our coordinates so that we are along  ${\vec r}=r{\hat z}$ direction, and the origin is at ${\vec x}_0$, so at large distances we have 
\beq
\vec{r}_1&=&r{\hat z}-\frac{1}{2}\rho{\hat d}= r {\hat z}\left(1- \frac{\rho}{2r} \cos\theta\right)-\frac{1}{2}\rho \sin\theta \left( {\hat x}  \cos\phi +{\hat y}  \sin \phi \right)~.
\eeq
\beq
|\vec{r}_1| &=&| r{\hat z}-\frac{1}{2}\rho{\hat d}| \approx \sqrt{r^2-r\rho \cos\theta} 
\approx r\left(1-\frac{\rho}{2r} \cos\theta\right)~.
\eeq
So the radial magnetic field is 
\beq
B^a_3 T^a&\approx & \frac{Q_1 }{g\,r^2} \left(1- \frac{\rho}{r} \cos\theta\right)   + \frac{Q_2 }{g\,r^2}\left(1+ \frac{\rho}{r} \cos\theta\right)  ~,
\eeq

The local kinetic terms for the massless gauge fields are
\beq
2{\rm Tr} F^a_{\mu \nu} T^a F^{b\mu \nu} T^b &=&2{\rm Tr} ( F_{(1)\mu \nu} Q_1 +F_{(2)\mu \nu} Q_2)(F_{(1)}^{\mu \nu} Q_1 +F_{(2)}^{\mu \nu} Q_2)\\
&=& F_{(1)\mu \nu}F_{(1)}^{\mu \nu}  +F_{(2)\mu \nu} F_{(2)}^{a\mu \nu}- F_{(1)\mu \nu}F_{(2)}^{\mu \nu} 
\eeq
Changing basis to the $Q$ and $X$ generators given in (\ref{diaggen}) and (\ref{brokengen}) we find:
\beq
A_Q^\mu Q+A_X^\mu X = A_{(1)}^\mu Q_1+A_{(2)}^\mu Q_2
\eeq
\beq
A_Q^\mu = \frac{1}{2}\left(A_{(1)}^\mu+A_{(2)}^\mu\right) ~,\quad\quad A_X^\mu = \frac{\sqrt{3}}{2}\left(A_{(1)}^\mu-A_{(2)}^\mu\right) ~.
\eeq
\beq
F_{(1)}^{\mu \nu}= F_{Q}^{\mu \nu}+\frac{1}{\sqrt{3}} F_{X}^{\mu \nu}~,\quad\quad F_{(2)}^{\mu \nu}= F_{Q}^{\mu \nu}-\frac{1}{\sqrt{3}} F_{X}^{\mu \nu}~.
\eeq
So the kinetic term becomes
\beq
F_{(1)\mu \nu}F_{(1)}^{\mu \nu}  +F_{(2)\mu \nu} F_{(2)}^{a\mu \nu}- F_{(1)\mu \nu}F_{(2)}^{\mu \nu} = F_{A\mu \nu}F_{A}^{\mu \nu}  +F_{X\mu \nu} F_{X}^{\mu \nu}
\eeq

In the $Q$-$X$ basis the magnetic field is 
\beq
\label{eq:QXbasis}
B^a_3 T^a&\approx & \frac{1 }{g\,r^2} \left(1- \frac{\rho}{r} \cos\theta\right) \left(Q+\frac{1}{\sqrt{3}}X\right)  + \frac{1 }{g\,r^2}\left(1+ \frac{\rho}{r} \cos\theta\right) \left(Q-\frac{1}{\sqrt{3}}X\right) \nonumber\\
&\approx & 2\frac{ Q }{g\,r^2}  + \frac{2}{\sqrt{3}}\, \rho \cos\theta \frac{X }{g\,r^3}~.
\eeq
while at ${\vec r}=0$ we have
\beq
B^a_3 T^a&=& \frac{-\frac{1}{2}\rho \cos \theta}{g|\frac{1}{2}\rho {\hat d}|^3} \left(Q+\frac{1}{\sqrt{3}}X\right)   + \frac{\frac{1}{2}\rho \cos \theta}{g|\frac{1}{2}\rho {\hat d}|^3} \left(Q-\frac{1}{\sqrt{3}}X\right)  \\
&=& \frac{-4 \cos \theta}{g \rho ^2} \left(Q+\frac{1}{\sqrt{3}}X\right)  \hat{r}^i  + \frac{4 \cos \theta}{g \rho^2} \left(Q-\frac{1}{\sqrt{3}}X\right)  \\
&=& \frac{-8 \cos \theta}{g \rho ^2} \frac{1}{\sqrt{3}}X~.
\eeq
Thus the $X$ gaugino component  of the zero mode is much more localized that the $A$ gaugino component.


\end{document}